 \documentclass[prb,preprint]{revtex4}

 \usepackage{dcolumn,amsmath}
 \usepackage{bm}
 \usepackage{graphicx} 

\newcommand{\tref}[1]{Table~\ref{#1}}

 \newcommand{\abinitio}{{\sl ab~initio} }

 \newcommand{\Cp}{Cn{}}

\begin{document}

\title
{
 Correlations beyond CCSD(T) for accurate study of Hg$_2$ and Cp$_2$
}

\author{A.N.\ Petrov}\email{anpetrov@pnpi.spb.ru}
                       \homepage{http://www.qchem.pnpi.spb.ru}
\altaffiliation [Also at ] {St.-Petersburg State University, St.-Petersburg,
        Russia}
\author{N.S.\ Mosyagin}
\author{A.V.\ Titov}

\affiliation{Petersburg Nuclear Physics Institute, 
             Gatchina, St.-Petersburg district 188300, Russia}

\author{A.V.\ Zaitsevskii}

\affiliation{HEPTI, RRC ``Kurchatov Institute'', 1 Kurchatov sq., 
 Moscow, 123182 Russia}

\author{E.A.\ Rykova}

\affiliation{Photochemistry Center, Russian Academy of Sciences, 
 Novatorov 7a, Moscow, 117421 Russia}

\date{\today}

\begin{abstract}
 Contributions from high-order (non-perturbative triple and quadruple)
 cluster amplitudes to the dissociation energies, equilibrium distances, and
 vibrational constants for the ground states of van der Waals dimers Hg$_2$
 and \Cp$_2$ are evaluated.  The incorporation of these contributions into the
 results of large-scale CCSD(T) calculations leads to non-negligible corrections
 of the computed molecular constants for Hg$_2$ (6\% for the dissociation
 energy), and enables one to attain perfect agreement with the experimental
 values.
\end{abstract}

\maketitle

\section{Introduction}

 The Hg dimer was the subject of a series of \abinitio calculations (see
 \tref{dataHg}). It is well understood from these studies that only
 highly-correlated relativistic calculations can give reliable results for such
 systems as the Hg$_2$ molecule in its ground state.  The widely used DFT method with
 the popular B88P86 and PW91 \cite{Perdew:92b} exchange-correlation functionals
 cannot ensure an acceptable accuracy of the description of the Hg\,--\,Hg bonding
 \cite{Anton:05}.
 {In turn, within the wavefunction-based approaches, the basis set superposition
 error (BSSE) can seriously deteriorate the results for such weakly bound
 molecules even for rather extensive basis sets.}
 To suppress the BSSE, large basis sets and counterpoise-like corrections should
 be employed.  Fully relativistic calculations appear to be computationally too
 demanding for highly-correlated treatments and can be performed only with
 unacceptable restrictions for the basis sets \cite{Nash:05}.  The calculated
 spectroscopic properties of Hg$_2$ reported in the papers \cite{Munro:01,
 Dolg:96} are in a pretty good agreement with the experimental data. In Ref.\
 \cite{Munro:01}, the scalar relativistic coupled cluster method with single,
 double and non-iterative triple cluster amplitudes (CCSD(T)) and the
 (9s,8p,7d,3f)/[7s,6p,4d,3f] basis set were used to correlate the outermost 24
 electrons ($5d$ and $6s$ shells).  Then the effects of correlations with the
 $5s$ and $5p$ shells and the basis set extension to higher angular momentum
 functions were estimated at the MP2 level of theory.  Spin-orbit interactions
 were neglected at that stage. In Ref.\cite{Dolg:96}, 24-electron CCSD(T)
 calculations with an uncontracted (9s,9p,8d,4f,2g) basis set were performed.  The
 correlation contributions from the $5s$ and $5p$ shells were estimated at the
 MP4 level of theory. Then the spin-orbit contribution was taken into account
 using MP2 calculations. CCSD(T) calculations in Ref.
 \cite{Schwerdtfeger:01b} with the larger basis set (11s,10p,9d,4f,3g,2h)
 correlating 40 electrons ($5s$, $5p$, $5d$, and $6s$  shells) and with the
 spin-orbit correction taken from Ref.\cite{Dolg:96} are in a worse agreement with
 the experimental data.  As has been noted in Ref.\ \cite{Schwerdtfeger:01b}, the
 good agreement of the calculated spectroscopic properties in papers
 \cite{Munro:01, Dolg:96} with the experimental data is a result of the
 fortunate cancellation of errors.  Our CCSD(T) calculations \cite{Petrov:09a}
 with the generalized correlation (14s,12p,9d,5f,3g,2h)/[10s,9p,7d,5f,3g,2h]
 basis set and spin-orbit contributions from the density-functional theory give
 results comparable with those in the work \cite{Schwerdtfeger:01b}.  We believe
 that the visible difference between our results and those obtained in Ref.\
 \cite{Schwerdtfeger:01b} can be attributed to different relativistic effective
 core potential models used in the calculations.

Recently long-lived isotopes of the heavier homologue of mercury, Copernicum
(\Cp\ or element 112), have been synthesized (see \cite{FLNRrep:08} and reference
therein).  At present, the experiments on \Cp\ are limited only by the
thermochromatography adsorption on the gold surface \cite{FLNRrep:08}; several
quantum chemical calculations on its compounds and, in particular,  \Cp$_2$
were performed to investigate the chemical properties of the new element (including inertness and
van der Waals radius).  The available theoretical data for the \Cp\ dimer (see
\tref{dataCp}) are rather contradictory. The difficulties in the
calculations on \Cp$_2$ are similar to those for Hg$_2$, except for the
dramatic increase of the relativistic effects (which, in turn, also affects the electron
correlations because of the
contraction of the $s$ and $p_{1/2}$ shells).
Since there are no experimental data on the spectroscopic constants of \Cp$_2$,
it was important to use the valid method tested in calculations on the Hg$_2$
dimer.

 The aim of the present work was to study the role of non-perturbative triple as
 well as of
 quadruple cluster amplitudes in the bonding of van der Waals dimers Hg$_2$
 and \Cp$_2$ for which accurate calculations of spectroscopic constants are
 required.
(By the effect of non-perturbative (or iterative) triples is meant  the
difference between CCSDT and CCSD(T), whereas by that of quadruples is meant
the difference between CCSDTQ and CCSDT.)
 The letter is of interest, first of all, from the viewpoint of extracting the van der Waals parameters
 (vdW radius etc.) of the \Cp\ atom.

 \section{Calculations and discussion}

 The contribution from the iteration of triple and quadruple cluster
 amplitudes for the four valence electrons in Hg$_2$ and \Cp$_2$ was evaluated
 as the difference between the total energies obtained in four-electron
 4e-FCI and 4e-CCSD(T) calculations \cite{Mosyagin:09aE}. 
   The account of similar contributions from the outer-core electrons is extremely
   expensive, but they are not expected to be noticeable and are not considered here.
  Scalar relativistic calculations were performed within the generalized
  relativistic effective core potential (GRECP) model \cite{Titov:99,
  Petrov:04b, Mosyagin:05a, Mosyagin:05b} using the {\sc MOLCAS} \cite{MOLCAS}
  code. Generalized correlation bases \cite{Mosyagin:00, Mosyagin:01b}
  comprising the $(14s,12p,9d,3f)/[7s,7p,3d,2f]$ and
  $(16s,21p,16d,12f,14g)/[4s,6p,3d,2f,1g]$ sets were used for Hg$_2$ and \Cp$_2$,
  respectively. These basis sets were chosen as a compromise between the accuracy
  and computational efforts in FCI calculations.  The accuracy was checked by
  calculating the contribution from non-iterative triple cluster
  amplitudes for the given and rather large
  $(14s,12p,9d,5f,3g,2h)/[10s,9p,7d,5f,3g,2h]$ and
  $(16s,21p,16d,12f,14g,1h)/[11s,10p,8d,5f,4g,1h]$ basis sets used in Ref.\
  \cite{Petrov:09a}.  These calculations were performed at the point $R=7.0$ a.u.\
  for Hg$_2$ and $R=6.5$ a.u.\ for
 \Cp$_2$ which are close to the equilibrium geometry.

  The corresponding
  contributions from the perturbative (non-iterative) triple cluster amplitudes
  are $-90$~cm$^{-1}$ (large basis set) and $-84$~cm$^{-1}$ (small basis set)
  for Hg$_2$, whereas they are $-51$~cm$^{-1}$ (large basis set) and
  $-43$~cm$^{-1}$ (small basis set) for \Cp$_2$.  Then the contributions from
  non-perturbed triples and quadruples for both dimers were evaluated for
  internuclear distances from 5 to 20 a.u.\ at a step of 1 a.u. and added
  to the best potential energy curves obtained in \cite{Petrov:09a} by the CCSD(T) method with large
  basis sets taking into account the counterpoise and spin-orbit
  corrections.
The spectroscopic constants derived from these curves are collected 
  in \tref{dataHg} and \tref{dataCp}.  One can see in \tref{dataHg}
  that the agreement with the experimental data is significantly improved.
 \tref{point} presents the energy differences  
$E(7.0\ {\rm a.u.})-E(\infty{})$ for Hg$_2$ and $E(6.5\ {\rm
  a.u.})-E(\infty{})$ for \Cp$_2$
(the chosen  finite distances are close to equilibrium values of the best
 theoretical level) calculated at the SCF,
 4e-CCSD, 4e-CCSD(T), 4e-CCSDT, and 4e-FCI levels of theory using small
 basis sets. One can deduce from these data that the total
 contributions from the connected quadruples and from the iteration of triples are
 $-23$~cm$^{-1}$ and $-6$~cm$^{-1}$ for Hg$_2$ and \Cp$_2$, respectively.  Since
 the contributions from only the iteration of triples are $-20$~cm$^{-1}$ and
 $-5$~cm$^{-1}$, we can conclude that the effect of the iterative accounting for the
 triple cluster amplitudes is noticeable, whereas the accounting for the quadruples is
 not.
 The potential energy curves for the Hg$_2$ and \Cp$_2$ dimers obtained in SCF calculations
 are repulsive. They remain repulsive also if only valence $s$ shells are
 correlated. An interesting feature that can be seen from the data in
 \tref{point} is that the potential energy curve of the \Cp$_2$ dimer becomes even more
 repulsive of only the {\it valence $7s$ shell is correlated}.
 To corroborate the reliability of the DFT-based spin-orbit corrections from Ref.\
 \cite{Petrov:09a} obtained with the generalized-gradient approximations for the
 exchange-correlation functional, i.e., Perdew-Wang (PW91) \cite{Perdew:92b} and
 Perdew-Burke-Ernzerhof (PBE) \cite{Perdew:96} models, we performed similar
 calculations with the meta-GGA TPSS functional \cite{Tao:03}. All three
 correcting functions were found to be nearly identical.

\section{Conclusions}

The effects of the non-perturbative treatment of triples and the inclusion of quadruples
on the computed Hg\,--\,Hg and \Cp\,--\,\Cp\ van der Waals interaction energies
are evaluated. For Hg$_2$, the account of these contributions substantially improves the CCSD(T) results, bringing the computed dissociation
energies, equilibrium distances, and vibrational constants into nearly perfect
agreement with the experimental values.  One can suppose that the non-perturbative
treatment of triples can play important role in the accurate description of other van
der Waals complexes.

\section*{ACKNOWLEDGMENTS}
 
 The present work was supported by RFBR, grants 07--03--01139 and 09--03--00655,
 and, partially, by the RFBR grant 09--03--01034. AP is grateful to the grant from
 the Ministry of Education and Science of the Russian Federation (Program for
 the Development of the Scientific Potential of Higher School, Grant No.\ 2.1.1/1136)


\clearpage



\begin{thebibliography}{20}
\expandafter\ifx\csname natexlab\endcsname\relax\def\natexlab#1{#1}\fi
\expandafter\ifx\csname bibnamefont\endcsname\relax
  \def\bibnamefont#1{#1}\fi
\expandafter\ifx\csname bibfnamefont\endcsname\relax
  \def\bibfnamefont#1{#1}\fi
\expandafter\ifx\csname citenamefont\endcsname\relax
  \def\citenamefont#1{#1}\fi
\expandafter\ifx\csname url\endcsname\relax
  \def\url#1{\texttt{#1}}\fi
\expandafter\ifx\csname urlprefix\endcsname\relax\def\urlprefix{URL }\fi
\providecommand{\bibinfo}[2]{#2}
\providecommand{\eprint}[2][]{\url{#2}}

\bibitem[{\citenamefont{Perdew et~al.}(1992)\citenamefont{Perdew, Chevary,
  Vosko, Jackson, Pederson, D.J.Singh, and C.Fiolhais}}]{Perdew:92b}
\bibinfo{author}{\bibfnamefont{J.~P.} \bibnamefont{Perdew}},
  \bibinfo{author}{\bibfnamefont{J.~A.} \bibnamefont{Chevary}},
  \bibinfo{author}{\bibfnamefont{S.~H.} \bibnamefont{Vosko}},
  \bibinfo{author}{\bibfnamefont{K.~A.} \bibnamefont{Jackson}},
  \bibinfo{author}{\bibfnamefont{M.~R.} \bibnamefont{Pederson}},
  \bibinfo{author}{\bibnamefont{D.J.Singh}}, \bibnamefont{and}
  \bibinfo{author}{\bibnamefont{C.Fiolhais}}, \bibinfo{journal}{Phys.\ Rev.\ B}
  \textbf{\bibinfo{volume}{46}}, \bibinfo{pages}{6671} (\bibinfo{year}{1992}).

\bibitem[{\citenamefont{Anton et~al.}(2005)\citenamefont{Anton, Fricke, and
  Schwerdtfeger}}]{Anton:05}
\bibinfo{author}{\bibfnamefont{J.}~\bibnamefont{Anton}},
  \bibinfo{author}{\bibfnamefont{B.}~\bibnamefont{Fricke}}, \bibnamefont{and}
  \bibinfo{author}{\bibfnamefont{P.}~\bibnamefont{Schwerdtfeger}},
  \bibinfo{journal}{Chem.\ Phys.} \textbf{\bibinfo{volume}{311}},
  \bibinfo{pages}{97} (\bibinfo{year}{2005}).

\bibitem[{\citenamefont{Nash}(2005)}]{Nash:05}
\bibinfo{author}{\bibfnamefont{C.~S.} \bibnamefont{Nash}},
  \bibinfo{journal}{J.\ Phys.\ Chem.\ A} \textbf{\bibinfo{volume}{109}},
  \bibinfo{pages}{3493} (\bibinfo{year}{2005}).

\bibitem[{\citenamefont{Munro et~al.}(2001)\citenamefont{Munro, Johnson, and
  Jordan}}]{Munro:01}
\bibinfo{author}{\bibfnamefont{L.~J.} \bibnamefont{Munro}},
  \bibinfo{author}{\bibfnamefont{J.~K.} \bibnamefont{Johnson}},
  \bibnamefont{and} \bibinfo{author}{\bibfnamefont{K.~D.}
  \bibnamefont{Jordan}}, \bibinfo{journal}{J.\ Chem.\ Phys.}
  \textbf{\bibinfo{volume}{114}}, \bibinfo{pages}{5545} (\bibinfo{year}{2001}).

\bibitem[{\citenamefont{Dolg and Flad}(1996)}]{Dolg:96}
\bibinfo{author}{\bibfnamefont{M.}~\bibnamefont{Dolg}} \bibnamefont{and}
  \bibinfo{author}{\bibfnamefont{H.-J.} \bibnamefont{Flad}},
  \bibinfo{journal}{J.\ Chem.\ Phys.} \textbf{\bibinfo{volume}{100}},
  \bibinfo{pages}{6147} (\bibinfo{year}{1996}).

\bibitem[{\citenamefont{Schwerdtfeger et~al.}(2001)\citenamefont{Schwerdtfeger,
  Wesendrup, Moyano, Sadlej, Greif, and Hensel}}]{Schwerdtfeger:01b}
\bibinfo{author}{\bibfnamefont{P.}~\bibnamefont{Schwerdtfeger}},
  \bibinfo{author}{\bibfnamefont{R.}~\bibnamefont{Wesendrup}},
  \bibinfo{author}{\bibfnamefont{G.~E.} \bibnamefont{Moyano}},
  \bibinfo{author}{\bibfnamefont{A.~J.} \bibnamefont{Sadlej}},
  \bibinfo{author}{\bibfnamefont{J.}~\bibnamefont{Greif}}, \bibnamefont{and}
  \bibinfo{author}{\bibfnamefont{F.}~\bibnamefont{Hensel}},
  \bibinfo{journal}{J.\ Chem.\ Phys.} \textbf{\bibinfo{volume}{115}},
  \bibinfo{pages}{7401} (\bibinfo{year}{2001}).

\bibitem[{\citenamefont{Petrov et~al.}(2009)\citenamefont{Petrov, Mosyagin,
  Titov, Zaitsevskii, and Rykova}}]{Petrov:09a}
\bibinfo{author}{\bibfnamefont{A.~N.} \bibnamefont{Petrov}},
  \bibinfo{author}{\bibfnamefont{N.~S.} \bibnamefont{Mosyagin}},
  \bibinfo{author}{\bibfnamefont{A.~V.} \bibnamefont{Titov}},
  \bibinfo{author}{\bibfnamefont{A.~V.} \bibnamefont{Zaitsevskii}},
  \bibnamefont{and} \bibinfo{author}{\bibfnamefont{E.~A.}
  \bibnamefont{Rykova}}, \bibinfo{journal}{Sov.\ J.\ Nucl.\ Phys.}
  \textbf{\bibinfo{volume}{72}}, \bibinfo{pages}{396} (\bibinfo{year}{2009}).

\bibitem[{FLN(2008)}]{FLNRrep:08}
in \emph{\bibinfo{booktitle}{JINR Annual Report 2008 (859.044)}}
  (\bibinfo{year}{2008}), pp. \bibinfo{pages}{86--96},
  \urlprefix\url{http://www1.jinr.ru/ Reports/2008/english/ 06\_flnr\_e.pdf}.

\bibitem[{\citenamefont{Mosyagin et~al.}(2009)\citenamefont{Mosyagin, Petrov,
  Titov, Zaitsevskii, and Rykova}}]{Mosyagin:09aE}
\bibinfo{author}{\bibfnamefont{N.~S.} \bibnamefont{Mosyagin}},
  \bibinfo{author}{\bibfnamefont{A.~N.} \bibnamefont{Petrov}},
  \bibinfo{author}{\bibfnamefont{A.~V.} \bibnamefont{Titov}},
  \bibinfo{author}{\bibfnamefont{A.~V.} \bibnamefont{Zaitsevskii}},
  \bibnamefont{and} \bibinfo{author}{\bibfnamefont{E.~A.} \bibnamefont{Rykova}}
  (\bibinfo{year}{2009}), \bibinfo{note}{[arXiv: 0901.0077]}.

\bibitem[{\citenamefont{Titov and Mosyagin}(1999)}]{Titov:99}
\bibinfo{author}{\bibfnamefont{A.~V.} \bibnamefont{Titov}} \bibnamefont{and}
  \bibinfo{author}{\bibfnamefont{N.~S.} \bibnamefont{Mosyagin}},
  \bibinfo{journal}{Int.\ J.\ Quantum Chem.} \textbf{\bibinfo{volume}{71}},
  \bibinfo{pages}{359} (\bibinfo{year}{1999}).

\bibitem[{\citenamefont{Petrov et~al.}(2004)\citenamefont{Petrov, Mosyagin,
  Titov, and Tupitsyn}}]{Petrov:04b}
\bibinfo{author}{\bibfnamefont{A.~N.} \bibnamefont{Petrov}},
  \bibinfo{author}{\bibfnamefont{N.~S.} \bibnamefont{Mosyagin}},
  \bibinfo{author}{\bibfnamefont{A.~V.} \bibnamefont{Titov}}, \bibnamefont{and}
  \bibinfo{author}{\bibfnamefont{I.~I.} \bibnamefont{Tupitsyn}},
  \bibinfo{journal}{J.\ Phys.\ B} \textbf{\bibinfo{volume}{37}},
  \bibinfo{pages}{4621} (\bibinfo{year}{2004}).

\bibitem[{\citenamefont{Mosyagin et~al.}(2006)\citenamefont{Mosyagin, Petrov,
  Titov, and Tupitsyn}}]{Mosyagin:05a}
\bibinfo{author}{\bibfnamefont{N.~S.} \bibnamefont{Mosyagin}},
  \bibinfo{author}{\bibfnamefont{A.~N.} \bibnamefont{Petrov}},
  \bibinfo{author}{\bibfnamefont{A.~V.} \bibnamefont{Titov}}, \bibnamefont{and}
  \bibinfo{author}{\bibfnamefont{I.~I.} \bibnamefont{Tupitsyn}}, in
  \emph{\bibinfo{booktitle}{Recent Advances in the Theory of Chemical and
  Physical Systems}}, edited by \bibinfo{editor}{\bibfnamefont{J.-P.}
  \bibnamefont{Julien}},
  \bibinfo{editor}{\bibfnamefont{J.}~\bibnamefont{Maruani}},
  \bibinfo{editor}{\bibfnamefont{D.}~\bibnamefont{Mayou}},
  \bibinfo{editor}{\bibfnamefont{S.}~\bibnamefont{Wilson}}, \bibnamefont{and}
  \bibinfo{editor}{\bibfnamefont{G.}~\bibnamefont{{Delgado-Barrio}}}
  (\bibinfo{publisher}{Springer}, \bibinfo{address}{Dordrecht, The
  Netherlands}, \bibinfo{year}{2006}), vol. \bibinfo{volume}{B~15} of
  \emph{\bibinfo{series}{Progr.\ Theor.\ Chem.\ Phys.}}, pp.
  \bibinfo{pages}{229--251}.

\bibitem[{\citenamefont{Mosyagin and Titov}(2005)}]{Mosyagin:05b}
\bibinfo{author}{\bibfnamefont{N.~S.} \bibnamefont{Mosyagin}} \bibnamefont{and}
  \bibinfo{author}{\bibfnamefont{A.~V.} \bibnamefont{Titov}},
  \bibinfo{journal}{J.\ Chem.\ Phys.} \textbf{\bibinfo{volume}{122}},
  \bibinfo{pages}{234106} (\bibinfo{year}{2005}).

\bibitem[{\citenamefont{Andersson et~al.}(1999)\citenamefont{Andersson,
  Blomberg, {F\"ulscher}, {Karlstr\"om}, Lindh, Malmqvist, {Neogr\'ady}, Olsen,
  Roos, Sadlej et~al.}}]{MOLCAS}
\bibinfo{author}{\bibfnamefont{K.}~\bibnamefont{Andersson}},
  \bibinfo{author}{\bibfnamefont{M.~R.~A.} \bibnamefont{Blomberg}},
  \bibinfo{author}{\bibfnamefont{M.~P.} \bibnamefont{{F\"ulscher}}},
  \bibinfo{author}{\bibfnamefont{G.}~\bibnamefont{{Karlstr\"om}}},
  \bibinfo{author}{\bibfnamefont{R.}~\bibnamefont{Lindh}},
  \bibinfo{author}{\bibfnamefont{P.-A.} \bibnamefont{Malmqvist}},
  \bibinfo{author}{\bibfnamefont{P.}~\bibnamefont{{Neogr\'ady}}},
  \bibinfo{author}{\bibfnamefont{J.}~\bibnamefont{Olsen}},
  \bibinfo{author}{\bibfnamefont{B.~O.} \bibnamefont{Roos}},
  \bibinfo{author}{\bibfnamefont{A.~J.} \bibnamefont{Sadlej}},
  \bibnamefont{et~al.} (\bibinfo{year}{1999}), \bibinfo{note}{quantum-chemical
  program package {``{\sc molcas}''}, Version 4.1}.

\bibitem[{\citenamefont{Mosyagin et~al.}(2000)\citenamefont{Mosyagin, Eliav,
  Titov, and Kaldor}}]{Mosyagin:00}
\bibinfo{author}{\bibfnamefont{N.~S.} \bibnamefont{Mosyagin}},
  \bibinfo{author}{\bibfnamefont{E.}~\bibnamefont{Eliav}},
  \bibinfo{author}{\bibfnamefont{A.~V.} \bibnamefont{Titov}}, \bibnamefont{and}
  \bibinfo{author}{\bibfnamefont{U.}~\bibnamefont{Kaldor}},
  \bibinfo{journal}{J.\ Phys.\ B} \textbf{\bibinfo{volume}{33}},
  \bibinfo{pages}{667} (\bibinfo{year}{2000}).

\bibitem[{\citenamefont{Mosyagin et~al.}(2001)\citenamefont{Mosyagin, Titov,
  Eliav, and Kaldor}}]{Mosyagin:01b}
\bibinfo{author}{\bibfnamefont{N.~S.} \bibnamefont{Mosyagin}},
  \bibinfo{author}{\bibfnamefont{A.~V.} \bibnamefont{Titov}},
  \bibinfo{author}{\bibfnamefont{E.}~\bibnamefont{Eliav}}, \bibnamefont{and}
  \bibinfo{author}{\bibfnamefont{U.}~\bibnamefont{Kaldor}},
  \bibinfo{journal}{J.\ Chem.\ Phys.} \textbf{\bibinfo{volume}{115}},
  \bibinfo{pages}{2007} (\bibinfo{year}{2001}).

\bibitem[{\citenamefont{Perdew et~al.}(1996)\citenamefont{Perdew, Burke, and
  Ernzerhof}}]{Perdew:96}
\bibinfo{author}{\bibfnamefont{J.~P.} \bibnamefont{Perdew}},
  \bibinfo{author}{\bibfnamefont{K.}~\bibnamefont{Burke}}, \bibnamefont{and}
  \bibinfo{author}{\bibfnamefont{M.}~\bibnamefont{Ernzerhof}},
  \bibinfo{journal}{Phys.\ Rev.\ Lett.} \textbf{\bibinfo{volume}{77}},
  \bibinfo{pages}{3865} (\bibinfo{year}{1996}).

\bibitem[{\citenamefont{Tao et~al.}(2003)\citenamefont{Tao, Perdew, Staroverov,
  and Scuseria}}]{Tao:03}
\bibinfo{author}{\bibfnamefont{J.}~\bibnamefont{Tao}},
  \bibinfo{author}{\bibfnamefont{J.~P.} \bibnamefont{Perdew}},
  \bibinfo{author}{\bibfnamefont{V.~N.} \bibnamefont{Staroverov}},
  \bibnamefont{and} \bibinfo{author}{\bibfnamefont{G.~E.}
  \bibnamefont{Scuseria}}, \bibinfo{journal}{Phys.\ Rev.\ Lett.}
  \textbf{\bibinfo{volume}{91}}, \bibinfo{pages}{146401}
  (\bibinfo{year}{2003}).

\bibitem[{\citenamefont{Motegi et~al.}(2001)\citenamefont{Motegi, Nakajima,
  Hirao, and Seijo}}]{Motegi:01}
\bibinfo{author}{\bibfnamefont{K.}~\bibnamefont{Motegi}},
  \bibinfo{author}{\bibfnamefont{T.}~\bibnamefont{Nakajima}},
  \bibinfo{author}{\bibfnamefont{K.}~\bibnamefont{Hirao}}, \bibnamefont{and}
  \bibinfo{author}{\bibfnamefont{L.}~\bibnamefont{Seijo}},
  \bibinfo{journal}{J.\ Chem.\ Phys.} \textbf{\bibinfo{volume}{114}},
  \bibinfo{pages}{6000} (\bibinfo{year}{2001}).

\bibitem[{\citenamefont{Koperski et~al.}(1994)\citenamefont{Koperski, Atkinson,
  and Krause}}]{Koperski:94}
\bibinfo{author}{\bibfnamefont{J.}~\bibnamefont{Koperski}},
  \bibinfo{author}{\bibfnamefont{J.~B.} \bibnamefont{Atkinson}},
  \bibnamefont{and} \bibinfo{author}{\bibfnamefont{L.}~\bibnamefont{Krause}},
  \bibinfo{journal}{Chem.\ Phys.\ Lett.} \textbf{\bibinfo{volume}{219}},
  \bibinfo{pages}{163} (\bibinfo{year}{1994}).

\end{thebibliography}

\begin{table}
\caption
{
Calculated equilibrium distance (in $\AA$) and 
other spectroscopic constants (in cm$^{-1}$) for the ground state
of the Hg$_2$ molecule as compared to the experimental and previously calculated data.
}
\vspace{0.5 cm}
\begin{ruledtabular}
\begin{tabular}{ccccc}
  method
  & $R_e$ & $D_e$ & $w_e$ & $w_ex_e$ \\
\hline
{ DFT(B88/P86)\cite{Anton:05}    } &   3.63    &   73         &   14           &       \\
{ DFT(PW91)\cite{Anton:05}      } &   3.55    &  385         &   24           &       \\
{ $^{a}$CCSD(T)\cite{Nash:05}      } &   3.60    &  581         &                &       \\
{ CCSD(T) $+\Delta E_{SO}$ \cite{Dolg:96}  } &   3.729   &  379         &   19.4         &  0.24 \\          
{ CCSD(T)  \cite{Munro:01}  }  &   3.718   &  379         &   19.4         &  0.24 \\
{ CCSD(T)  \cite{Motegi:01} } &   3.836   &  315         &   16.3         &       \\
{  CCSD(T) $+\Delta E_{SO}$ \cite{Schwerdtfeger:01b} } & 3.743 & 328 & 18.4 & 0.28 \\
{ CCSD(T) $+\Delta E_{SO}$ \cite{Petrov:09a}} &  { 3.730 } &  { 355  } &  { 18.74 } &  {  0.24  } \\
{ This work
} 
&  { 3.711 } &  { 377  } &  { 19.32 } &  {  0.24  } \\
 { Experiment \cite{Koperski:94} }  &  { $3.69 \pm 0.01$ } & { $380 \pm 25$ } & { $19.6 \pm 0.3$} & { $0.25 \pm 0.05$}
\end{tabular}
\end{ruledtabular}
\begin{flushleft}
$^{\rm a}$ Total relativistic version of the coupled-cluster
method. Small basis set, BSSE is not compensated.
\end{flushleft}
\label{dataHg}
\end{table}

\begin{table}
\caption
{
 Calculated equilibrium distance (in $\AA$) and 
other spectroscopic constants (in cm$^{-1}$) for the ground state
 of the \Cp$_2$ molecule as compared to the previously calculated data.}      
\vspace{0.5 cm}
\begin{ruledtabular}
\begin{tabular}{ccccc}
  method
  & $R_e$ & $D_e$ & $w_e$ & $w_ex_e$ \\
\hline
{ DFT(B88/P86)\cite{Anton:05}    } &  3.45     &  315         &   25           &  ???  \\
{ DFT(PW91)\cite{Anton:05}      } &   3.39    &  649         &   30           &  ???  \\
{ $^{a}$CCSD(T)\cite{Nash:05}      } & 3.07      & 1508         &                &       \\
{ CCSD(T) $+\Delta E_{SO}$ \cite{Petrov:09a}} &  { 3.323 } &  { 768  } &  { 29.83 } &  { 0.3272 } \\
{ This work } &  { 3.322 } &  { 775  } &  { 29.94 } & 
{  0.3266} \\
\end{tabular}
\end{ruledtabular}
\begin{flushleft}
$^{\rm a}$ Total relativistic version of the coupled-cluster
method. Small basis set, BSSE is not compensated.
\end{flushleft}
\label{dataCp}
\end{table}

\begin{table}
\caption
{
 Calculated energy differences (in cm$^{-1}$) $E(R=7.0~{\rm a.u.})-E(R=200.0~{\rm a.u.})$
 for Hg$_2$ and $E(R=6.5~{\rm a.u.})-E(R=200.0~{\rm a.u.})$  for
 \Cp$_2$}      
\vspace{0.5 cm}
\begin{ruledtabular}
\begin{tabular}{ldd}
               & \rm Hg_2  &  \rm \Cp_2  \\
\hline
    SCF        & 617.4   &    944.5  \\
    4e-CCSD    & 211.9   &   1573.3  \\
    4e-CCSD(T) & 128.2   &   1530.0  \\
    4e-CCSDT   & 108.3   &   1524.7  \\
    4e-FCI     & 105.4   &   1523.9
\end{tabular}
\end{ruledtabular}
\label{point}
\end{table}

\newcommand{\noopsort}[1]{} \newcommand{\printfirst}[2]{#1}
  \newcommand{\singleletter}[1]{#1} \newcommand{\switchargs}[2]{#2#1}

\end{document}